# GENERALIZED GAUSSIAN EFFECTIVE POTENTIAL :
# THERMAL CORRECTIONS

Paolo Cea    and    Luigi Tedesco

*Dipartimento di Fisica dell'Universitá di Bari, I-70126 Bari, Italy*

*INFN, Sezione di Bari, I-70126 Bari, Italy*

## ABSTRACT

We evaluate the thermal corrections to the generalized Gaussian effective potential. We carry out the calculations of the lowest order corrections in the case of self-interacting scalar fields in one and two spatial dimensions, and study the restoration of the symmetry at high temperatures.

In a previous study [1,2] we introduced the generalized Gaussian effective potential which allows to evaluate in a systematic manner the corrections to the Gaussian approximation. In this paper we would like to study the thermal corrections to the generalized Gaussian effective potential. In particular we are interested in the phenomenon of the symmetry restoration at high temperatures. To this end we shall discuss in details the lowest order thermal corrections for self-interacting scalar fields in one and two spatial dimensions.

Let us firstly discuss the thermal corrections in the general setting of self-interacting scalar fields in $\nu + 1$ dimensions. In the fixed-time Schrödinger representation, the Hamiltonian reads

$$H = \int d^\nu x \left[ -\frac{1}{2} \frac{\delta^2}{\delta \phi(\vec{x}) \delta \phi(\vec{x})} + \frac{1}{2} \left( \vec{\nabla} \phi(\vec{x}) \right)^2 + \frac{1}{2} m^2 \phi^2(\vec{x}) + \frac{\lambda}{4!} \phi^4(\vec{x}) \right]. \quad (1)$$

As it is well known [3], the Gaussian effective potential is given by

$$V_{GEP}(\phi_0) = \frac{1}{V} min_{|\psi_0>} \frac{<\psi_0|H|\psi_0>}{<\psi_0|\psi_0>}, \quad (2)$$

where $V$ is the spatial volume and $|\psi_0>$ is a Gaussian trial functional centered at $\phi_0$.

The generalized Gaussian effective potential is set up as follows. Starting from the vacuum functional $|\psi_0>$, one constructs a variational Fock basis $|n>$. After that, the Hamiltonian $H$ is splitted into the unperturbated Hamiltonian $H_0$ and the perturbation $H_I$. $H_0$ and $H_I$ are defined respectively by the diagonal and off-diagonal elements of $H$ on the basis $|n>$. By switching the interaction $H_I$ on, the unperturbed ground state $|\psi_0>$ evolves into $|\Omega>$. Whereupon the generalized Gaussian effective potential is given by [1]:

$$V_G(\phi_0) = \frac{1}{V} \frac{<\Omega|H|\Omega>}{<\Omega|\Omega>}, \quad (3)$$

with the constraint

$$\frac{<\Omega|\phi(\vec{x})|\Omega>}{<\Omega|\Omega>} = \phi_0. \tag{4}$$

Using the Gell-Mann and Low theorem on the ground state [4], we showed that

$$V_G(\phi_0) = V_{GEP}(\phi_0) + \frac{1}{V}\sum_{n=1}^{\infty}\frac{(-i)^n}{n!}\int_{-\infty}^{0}dt_1...\int_{-\infty}^{0}dt_n\cdot$$
$$\cdot <0|T(H_I(0)H_I(t_1)...H_I(t_n))|0>_{connected} e^{[\epsilon(t_1+...+t_n)]}, \tag{5}$$

where $|0>=|\psi_0>$, and $H_I(t)$ is the interaction Hamiltonian in the interaction representation.

We, now, turn to finite-temperature field theory. Thermal equilibrium is achieved when the system has minimized its free energy F. Standard thermodinamics [5] gives

$$F = -\frac{1}{\beta}lnZ, \tag{6}$$

where $Z$ is the partition function

$$Z = tr\left(e^{-\beta H}\right) \tag{7}$$

and $\beta = \frac{1}{T}$ (Boltzmann's constant $k=1$).

The finite temperature generalized Gaussian effective potential is defined by

$$V_G^T(\phi_0) = \frac{F}{V} = -\frac{1}{\beta V}lnZ \tag{8}$$

with the constraint (4).

In order to evaluate the partition function we face with the problem of calculating the trace in Eq. (7). Indeed, the trace involves a summation over the unknown eigenstates of the full Hamiltonian $H$. However, in our approach the full Hamiltonian has been splitted into the unperturbated Hamiltonian $H_0$ and the perturbation $H_I$. Thus the natural strategy is to perform the trace in Eq. (7) by employing the thermodinamic perturbation theory [6]. Indeed, to first order in $H_I$, the trace can be replaced by a summation over the eigenstates of $H_0$. A straightforward calculation gives:

$$Z = Z_0(1 - \beta <H_I>^\beta), \tag{9}$$

where

$$Z_0 = tr(e^{-\beta H_0}) = \sum_n <n|e^{-\beta H_0}|n> \tag{10}$$

and

$$<H_I>^\beta \equiv \frac{1}{Z_0} tr(e^{-\beta H_0} H_I) = \frac{1}{Z_0} \sum_n <n|e^{-\beta H_0} H_I|n>. \tag{11}$$

In Equation (10) and (11) we sum over the eigenstates of free scalar field Hamiltonian with mass $\mu$, where $\mu$ satisfies the gap equation:

$$\mu^2 = m^2 + \frac{\lambda}{2}\phi_0^2 + \frac{\lambda}{4}\int \frac{d^\nu k}{(2\pi)^\nu} \frac{1}{\sqrt{\vec{k} + \mu^2}}. \tag{12}$$

Observing that

$$<n|H_I|n> = 0, \tag{13}$$

and that the states $|n>$ are eigenstates of $H_0$, one readily gets

$$<H_I>^\beta = 0. \tag{14}$$

As a consequence, a straightforward calculation gives:

$$V_G^T(\phi_0) = V_{GEP}(\phi_0) + \frac{1}{\beta} \int \frac{d^\nu k}{(2\pi)^\nu} ln\left(1 - e^{-\beta g(\vec{k})}\right) + O[(H_I)^2], \qquad (15)$$

where $g(\vec{k}) = \sqrt{\vec{k}^2 + \mu^2}$. Note that Eq. (15) differs from the finite-temperature Gaussian effective potential [7]. The difference resides in the different use of the gap equation. In our scheme the gap equation (12) is fixed once and for all. In particular it does not depend on the temperature. On the other hand, in the finite-temperature Gaussian effective potential approach the gap equation includes the thermal corrections. Now, the gap equation fixes the basis to sum over in the thermal average. This means that different gap equations lead to inequivalent basis. As a matter of fact the discrepancy between our result Eq. (15) and the finite-temperature Gaussian effective potential comes from the thermal average of the interaction Hamiltonian. In our approach Eq.(14) holds, whereas in Ref. [7] $<H_I>^\beta \neq 0$.

As we have discussed in Ref. [1,2] the main advantages of our approach at zero temperature reside in the diagrammatic expansion of the higher order corrections which is amenable to a diagrammatic resummation. Moreover the renormalization of the generalized Gaussian effective potential relies on the underlying $\phi_0 = 0$ field theory. Remarkably enough, it turns out that these fine features extend also to the case of scalar field theory at finite temperatures. Indeed we can write a manageable formula for the higher order corrections to Eq. (15). To do this, the usual thermodinamic perturbation theory is useless. Instead we may follow the Matsubara's methods [8]. In the Matsubara's scheme one deals with scalar fields which depend on the fictious imaginary time $\tau$ varying in the interval $(0, \beta)$.

If the Hamiltonian of the system in thermal equilibrium can be written as $H = H_0 + H_I$, then one can show that the corrections to the thermodynamic potential are given by

$$\Delta\Omega = -\frac{1}{\beta} ln < T_\tau exp\left(-\int_0^\beta H_I(\tau)d\tau\right) >, \qquad (16)$$

where $H_I(\tau)$ is the interaction Hamiltonian in the Matsubara's interaction representation, $T_\tau$ is the $\tau$-ordering operator, and the thermal averages are done with the free field partition function. Moreover, it turns out that only connected diagrams contribute to $\Delta\Omega$:

$$\Delta\Omega = -\frac{1}{\beta}\sum_{m=1}^{\infty}\frac{(-1)^m}{m!}\int_0^\beta d\tau_1...\tau_m < T_\tau(H_I(\tau_1)...H_I(\tau_m)) >_{connected}^\beta. \qquad (17)$$

In our case, if we write

$$V_G^T(\phi_0) = V_{GEP}(\phi_0) + \frac{1}{\beta}\int\frac{d^\nu k}{(2\pi)^\nu}ln(1-e^{-\beta g(\vec{k})}) + \Delta V_G^T(\phi_0), \qquad (18)$$

we readily get

$$\Delta V_G(\phi_0) = -\frac{1}{\beta}\sum_{m=2}^{\infty}\frac{(-1)^m}{m!}\int_0^\beta d\tau_1...\tau_m < T_\tau(H_I(\tau_1)...H_I(\tau_m)) >_{connected}^\beta. \qquad (19)$$

Note that, due to Eq. (14), the sum in Eq. (19) starts from $m = 2$. The thermal average of time-ordered products is evaluated by means of the Wick's theorem for thermal fields [9]. In this way we obtain the thermal corrections to the generalized Gaussian effective potential as connected vacuum diagrams. In Figure 1 we display the second order thermal corrections. The vertices can be extracted from the interaction Hamiltonian [1,2]:

$$H_I = \int d^\nu x \left[\left(\mu^2\phi_0 - \frac{\lambda}{3}\phi_0^3\right) : \eta(\vec{x}) : +\frac{\lambda}{3!}\phi_0 : \eta^3(\vec{x}) : +\frac{\lambda}{4!} : \eta^4(\vec{x}) :\right], \qquad (20)$$

where $\eta(x) = \phi(x) - \phi_0$.

The solid lines in Fig.1 are the thermal propagators of the scalar fields with mass $\mu$:

$$G_\beta(\vec{x}-\vec{y}, \tau_1-\tau_2) = <T_\tau \eta(\vec{x},\tau_1)\eta(\vec{y},\tau_2)>^\beta = \frac{1}{\beta} \sum_{n=-\infty}^{+\infty} \int \frac{d^\nu k}{(2\pi)^\nu} \frac{e^{i[\vec{k}(\vec{x}-\vec{y})-\omega_n(\tau_1-\tau_2)]}}{\omega_n^2 + g^2(\vec{k})}, \quad (21)$$

where $\omega_n = \frac{2\pi n}{\beta}$. Note that the thermal propagator is periodic in the time variable with period $\frac{2\pi}{\beta}$.

A distinguishing feature of the graphical expansion of Eq.(19) with respect to Eq.(5) stems from the fact that the $(m!)^{-1}$ coming from the $mth$ order term is not completely cancelled by the number of different Wick contractions corresponding to a given graph. Consequently, a graph contributes to $\Delta V_G^T(\phi_0)$ in proportion to a combinatoric coefficient depending on the order of the graph. Moreover in evaluating the contribution due to a given graph one should take care of the normal ordering prescription in the interaction Hamiltonian. It turns out that the normal ordering in $H_I$ modifies the so-called anomalous diagrams, i.e. the diagrams which vanish at zero temperature [10]. For istance, in Fig. 1 the diagrams (b) and (c) are anomalous. We find that the normal ordering of the interaction Hamiltonian modifies the diagrams like (b) and (c) in Fig. 1 by replacing $G_\beta(0)$ with $\tilde{G}_\beta(0)$, where

$$\tilde{G}_\beta(0) = G_\beta(0) - \lim_{\beta \to \infty} G_\beta(0) = \int \frac{d^\nu k}{(2\pi)^\nu} \frac{1}{2g(\vec{k})} \left[ cotgh\left(\frac{\beta g(\vec{k})}{2}\right) - 1 \right]. \quad (22)$$

A full account of the second order thermal corrections to the generalized Gaussian effective potential will be presented elsewhere [11]. In the remainder of the present paper we would like to analyze the lowest order thermal corrections, Eq.(15), for the case of $\nu = 1, 2$ spatial dimensions.

In one spatial dimension Eq.(15) reads

$$V_G^T(\phi_0) = V_{GEP}(\phi_0) + \frac{\mu_0^2}{\pi \hat{\beta}^2} \int_0^\infty dt \, ln\left(1 - e^{-\sqrt{t^2 + \hat{\beta}^2 x}}\right), \qquad (23)$$

where $\hat{\beta} = \beta \mu_0$, $x = \frac{\mu^2}{\mu_0^2}$, and $\mu_0 = \mu(\phi_0 = 0)$. In Figure 2 we show $V_G^T(\phi_0) - V_G^T(0)$ (in units of $\mu_0^2$) versus $\phi_0$ for $\hat{\lambda} > \hat{\lambda}_c$ [*]. As we can see, the symmetry broken at T=0 gets restored for $\hat{T} > \hat{T}_c$. Obviously the critical temperature depends on $\hat{\lambda}$. It turns out that $\hat{T}_c$ can be extimate, within a few percent, by means of the high-temperature expansion of the integral in Eq. (23). By following the classical work by Dolan and Jackiw [12] we find the following high-temperature expansion:

$$V_G^T(\phi_0) = V_{GEP}(\phi_0) + \mu_0^2 \left[ -\frac{\pi}{2\hat{\beta}^2} + \frac{\sqrt{x}}{2\hat{\beta}} - \frac{x}{8\pi} + \frac{x}{4\pi} ln\left(\frac{\hat{\beta}\sqrt{x}}{4\pi}\right) - \frac{\zeta(3)}{64\pi^3} x^2 \hat{\beta}^2 + \right.$$

$$\left. \frac{\zeta(5)}{512\pi^5} \hat{\beta}^4 x^3 + O(\hat{\beta}^6) \right] \qquad (24)$$

where $\zeta(z)$ is the Riemann's zeta function.

In Fig.2 we also show the high-temperature expansion Eq. (24) (dashed line); we can see that the high-temperature expansion is a good approximation even at $T \simeq T_c$. Indeed For $\hat{\lambda} = 4$ we find $\hat{T}_c \simeq 1.27$, which agrees with the value obtained by using the high-temperature expansion (24).

The case of two spatial dimensions can be dealed with in a similar way. We have

$$V_G^T(\Phi_0) = V_{GEP}(\Phi_0) + \frac{\mu_0^3}{(2\pi)\hat{\beta}^3} \int_0^\infty dt \, t \, ln\left(1 - e^{-\sqrt{t^2 + \hat{\beta}^2 x}}\right), \qquad (25)$$

where now $\Phi_0 = \frac{\phi_0^2}{\mu_0}$.

The high temperature expansion of the integral in (25) gives [12]

---

[*] Notations as in Ref. [2]

$$V_G^T(\Phi_0) = V_{GEP}(\Phi_0) + \mu_0^3 \left[ -\frac{\pi^2}{90\hat{\beta}^4} + \frac{x}{24\hat{\beta}^2} - \frac{1}{12\pi}\frac{x^{3/2}}{\hat{\beta}} - \frac{x^2}{64\pi^2}ln(x\hat{\beta}^2) + \right.$$
$$\left. \frac{c}{64\pi}x^2 + O(\hat{\beta}^2) \right] \quad (26)$$

where the constant $c = \frac{3}{2} + 2ln4\pi - 2\gamma$, and $\gamma$ is Euler-Mascheroni's constant. In Figure 3 we display Eq. (25) together with Eq. (26) in units of $\mu_0^3$ (we subtract the temperature dependent constant $V_{\hat{\beta}}^T(0)$). Again, the thermal corrections lead to the expected symmetry restoration at high temperatures. In this case, however, the high-temperature expansion Eq.(26) is a valid approximation for $T >> T_c$ only. So that the critical temperature $T_c$ must be extimate by numerical methods.

In summary, we have discussed the thermal corrections to the generalized Gaussian effective potential. In the lowest order approximation we obtain the restoration of the symmetry at high temperatures by a first order phase transition. The second order thermal corrections are under study. We plan to present the results in a future publication [11].

**FIGURE CAPTIONS**

**Fig. 1** Second order thermal corrections to the generalized Gaussian effective potential.

**Fig.2** Thermal corrections to generalized Gaussian effective potential to the lowest order for $\nu = 1$ and $\hat{\lambda} = 4$. Dashed lines refer to high-temperature expansion.

**Fig.3** Thermal corrections to generalized Gaussian effective potential to the lowest order for two spatial dimensions and $\hat{\lambda} = 4$. Dashed line refers to high-temperature expansion. The critical temperature is $T_c \simeq 1.60$.

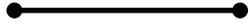

(a)

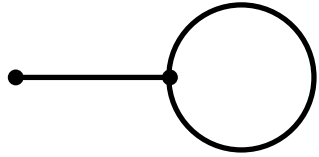

(b)

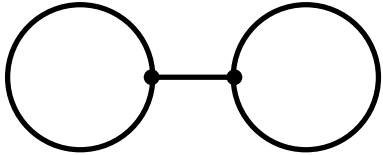

(c)

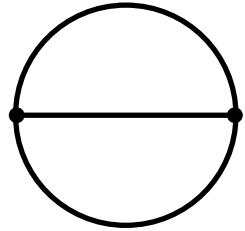

(d)

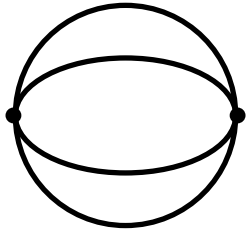

(e)

FIG. 1

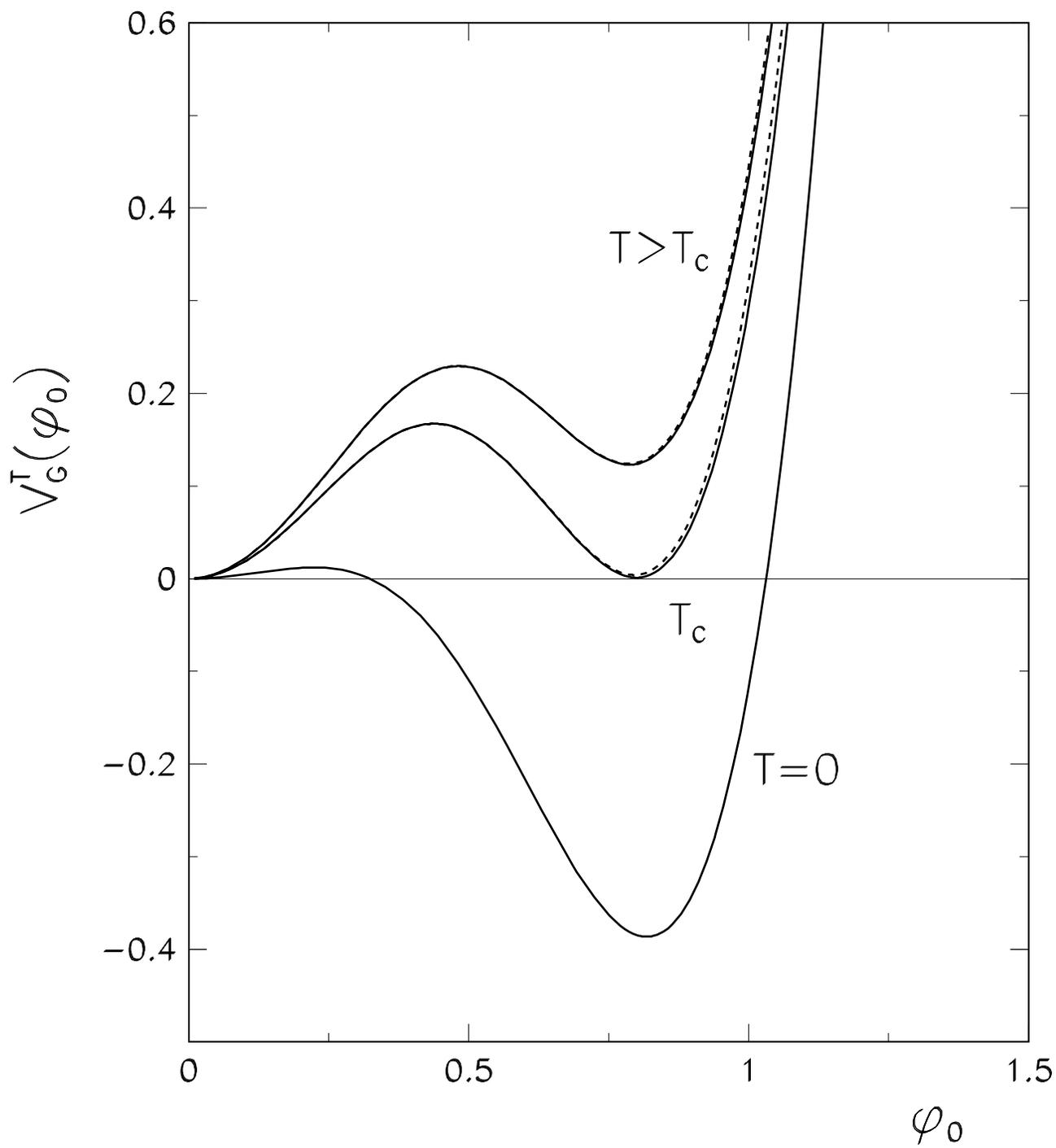

FIG. 2

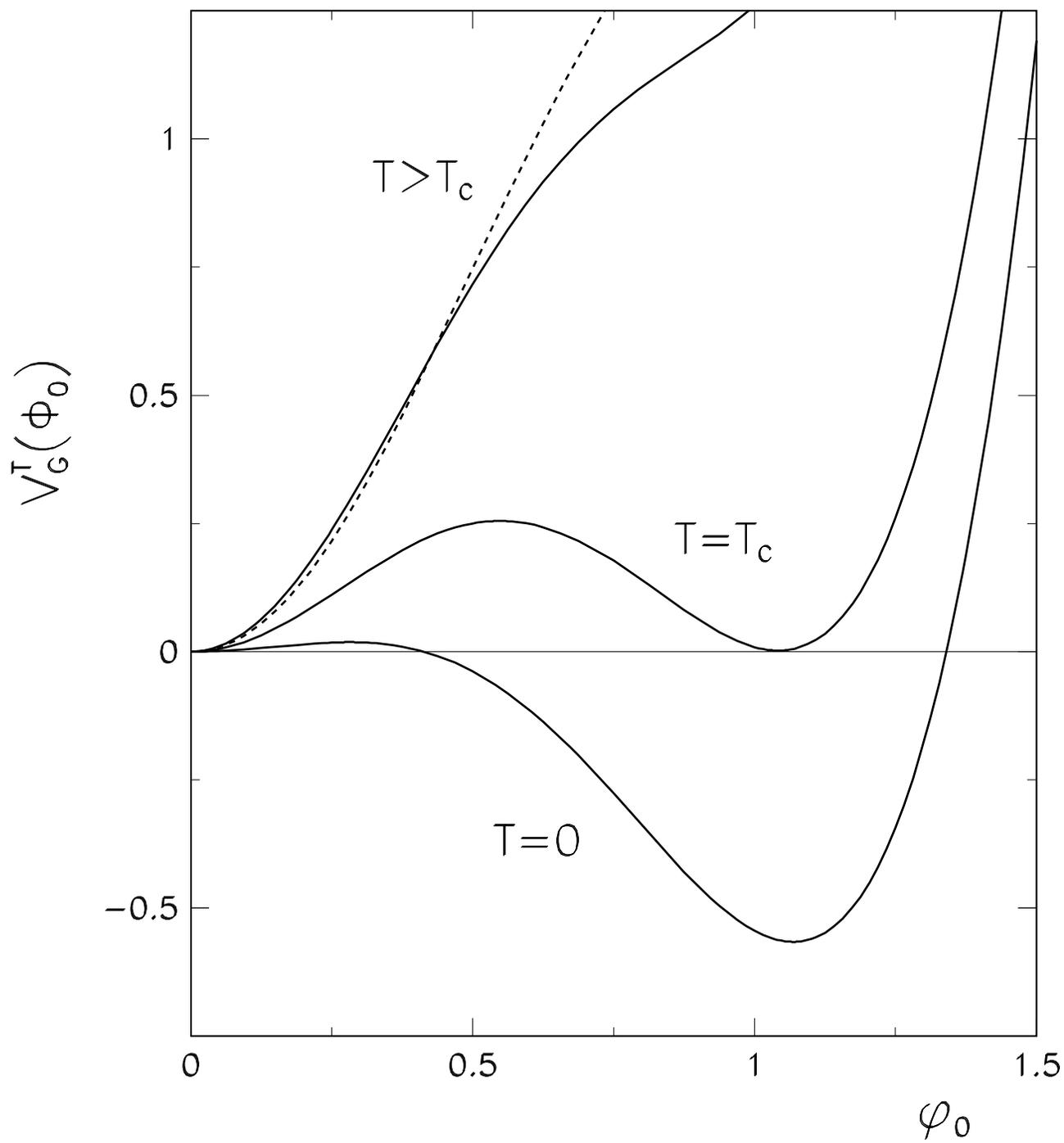

FIG. 3